\documentclass[]{assets/matterlab}
\usepackage{tabularx}
\usepackage{rotating}
\usepackage[normalem]{ulem}
\useunder{\uline}{\ul}{}
\newcommand{\sysname}{\textsc{TreeReader}}
\newcommand{\papertitle}{\sysname{}: A Hierarchical Academic Paper Reader\newline Powered by Language Models}

\renewcommand{\cite}{\citep}
\usepackage{microtype}
\usepackage{graphicx}
\usepackage{booktabs}
\usepackage{float}
\usepackage{xurl}
\usepackage{hyperref}
\usepackage{titlesec}

\usepackage{amsmath}
\usepackage{amssymb}
\usepackage{mathtools}
\usepackage{amsthm}
\usepackage{bm}

\usepackage[noabbrev,nameinlink]{cleveref}

\usepackage{amsmath,amsfonts,bm}

\def\eqref#1{equation~\ref{#1}}
\def\1{\bm{1}}

\DeclareMathAlphabet{\mathsfit}{\encodingdefault}{\sfdefault}{m}{sl}
\SetMathAlphabet{\mathsfit}{bold}{\encodingdefault}{\sfdefault}{bx}{n}

\usepackage{orcidlink}
\usepackage{comment}
\usepackage[version=4]{mhchem}
\usepackage{longtable}
\usepackage{array}
\renewcommand{\arraystretch}{1.7}
\usepackage{multirow}           % tabular cells spanning multiple rows
\usepackage{amsfonts}           % blackboard math symbols
\usepackage{nicefrac}
\usepackage{duckuments}         % sample images
\usepackage{thmtools,thm-restate}
\usepackage{enumitem}
\usepackage{xcolor,colortbl}
\usepackage{tikz}
\usepackage{tikz-cd}
\usepackage{caption}
\usepackage{subcaption}
\usepackage{listings}
\usepackage{gensymb}
\usepackage{textcomp} % Alternative to gensymb
\usepackage[inkscapelatex=false]{svg}

\newcommand{\addressCHEM}{Department of Chemistry, University of Toronto, Lash Miller Chemical Laboratories, 80 St. George Street, ON M5S 3H6, Toronto, Canada}
\newcommand{\addressAC}{Acceleration Consortium, 700 University Ave., M7A 2S4, Toronto, Canada}
\newcommand{\addressCS}{Department of Computer Science, University of Toronto, Sandford Fleming Building, 10 King’s College Road, ON M5S 3G4, Toronto, Canada}
\newcommand{\addressVECTOR}{Vector Institute for Artificial Intelligence, 661 University Ave. Suite 710, ON M5G 1M1, Toronto, Canada}
\newcommand{\addressMSE}{Department of Materials Science \& Engineering, University of Toronto, 184 College St., M5S 3E4, Toronto, Canada}
\newcommand{\addressCHEMENG}{Department of Chemical Engineering \& Applied Chemistry, University of Toronto, 200 College St. ON M5S 3E5, Toronto, Canada}
\newcommand{\addressCIFAR}{Senior Fellow, Canadian Institute for Advanced Research (CIFAR), 661 University Ave., M5G 1M1, Toronto, Canada}
\newcommand{\addressNVIDIA}{NVIDIA, 431 King St W \#6th, M5V 1K4, Toronto, Canada}

\newcommand{\addressUTM}{Department of Mathematical and Computational Sciences, University of Toronto Mississauga, 3359 Mississauga Road, Deerfield Hall, ON L5L 1C6, Mississauga, Canada}

\providecommand{\doi}[1]{\href{https://doi.org/#1}{doi:#1}}

\title{\papertitle}
\author[1,2]{Zijian Zhang}
\author[1,2]{Pan Chen}
\author[1]{Fangshi Du}
\author[1]{Runlong Ye}
\author[1]{Oliver Huang}
\author[4]{Michael Liut}
\author[1,2,3,5,6,7,8,9,*]{Al\'an Aspuru-Guzik}

\affiliation[1]{\addressCS}
\affiliation[2]{\addressVECTOR}
\affiliation[3]{\addressCHEM}
\affiliation[4]{\addressUTM}
\affiliation[5]{\addressMSE}
\affiliation[6]{\addressCHEMENG}
\affiliation[7]{\addressAC}
\affiliation[8]{\addressCIFAR}
\affiliation[9]{\addressNVIDIA}

\abstract{
Efficiently navigating and understanding academic papers is crucial for scientific progress. Traditional linear formats like PDF and HTML can cause cognitive overload and obscure a paper's hierarchical structure, making it difficult to locate key information. While LLM-based chatbots offer summarization, they often lack nuanced understanding of specific sections, may produce unreliable information, and typically discard the document's navigational structure. Drawing insights from a formative study on academic reading practices, we introduce \sysname{}, a novel language model-augmented paper reader. \sysname{} decomposes papers into an interactive tree structure where each section is initially represented by an LLM-generated concise summary, with underlying details accessible on demand. This design allows users to quickly grasp core ideas, selectively explore sections of interest, and verify summaries against the source text. A user study was conducted to evaluate \sysname{}'s impact on reading efficiency and comprehension. \sysname{} provides a more focused and efficient way to navigate and understand complex academic literature by bridging hierarchical summarization with interactive exploration.
}

\date{\today}
\correspondence{Al\'an Aspuru-Guzik at \email{aspuru@nvidia.com}}

\metadata[Code]{\url{https://github.com/aspuru-guzik-group/TreeReader}}

\begin{document}

\maketitle

\section{Introduction}

  \begin{center}
    \includegraphics[width=1.0\textwidth]{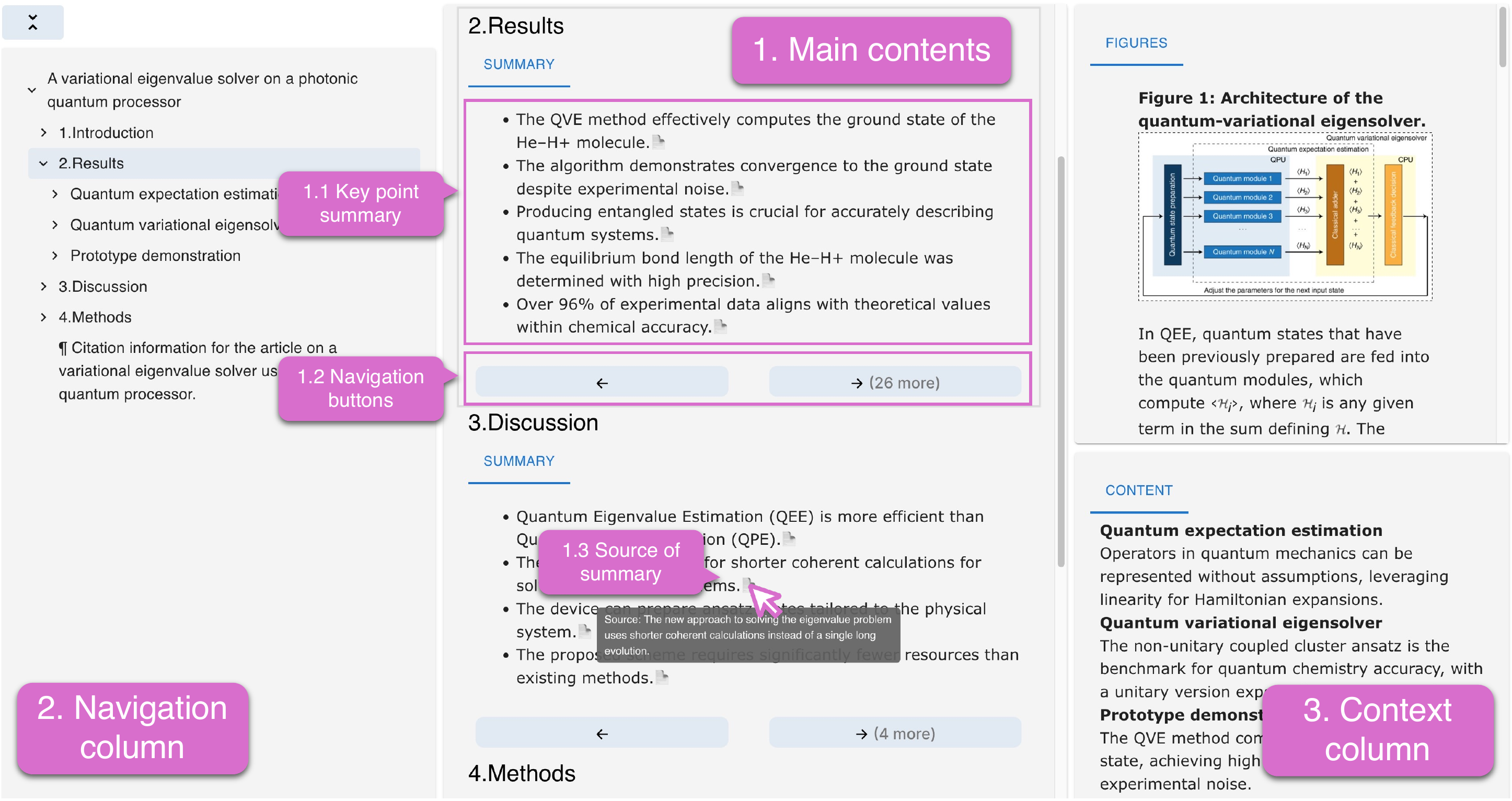}
    \captionof{figure}{User interface of \sysname{}. 1. (Main contents) \sysname{} presents subsections and paragraphs in a section by their key point summaries (see 1.1) instead of linearly displaying all their sub-content. The user can view the details of a subsection on demand or view the summary of the whole section by clicking ``→'' or ``←'' in the Navigation buttons (see 1.2). 
    The users can also check the quality of the summary by viewing their source texts on demand (see 1.3).
    2. (Navigation tree) The users can also quickly view and navigate to a certain section or paragraph through an interactive tree view on the left. 3. (Contextual information) When the user focuses on a subsection or paragraph in the Main contents, \sysname{} will display additional information, such as the figures in the section.}
    \label{fig:system}
    \vspace{1em}
  \end{center}

Academic publications (papers) are the primary medium for communicating scholarly knowledge, making their reading an essential, yet often time-consuming, part of scientific research across all disciplines.
With the ever-increasing volume and complexity of scientific literature, the ability to efficiently navigate and comprehend these documents is paramount for accelerating scientific progress. However, the predominant formats for academic papers, PDF and standard HTML, present content in a linear fashion, where each section is displayed in its entirety one after another. This linear presentation, despite papers typically adhering to an inherent hierarchical section structure, poses significant challenges for the users.

The undifferentiated stream of linear content can lead to cognitive overload. Extensive cognitive research demonstrates that human working memory is limited, processing only a few pieces of information effectively at once \cite{miller1956magical}, and that performance improves when information is grouped into meaningful ``chunks'' \cite{Thalmann:2019aa}. 
Linear formats, by presenting all details irrespective of their immediate relevance to the user's goals, make it difficult to discern these chunks or quickly grasp the core arguments. Consequently, users must often manually skip over content to find key information, a process that is both inefficient and prone to missing crucial details.
Experiments have shown that reorganizing material into an explicit outline yields significantly higher recall and comprehension \cite{McKoon1977OrganizationOI, doi:10.1177/21582440231191795}, highlighting the measurable cognitive benefits of hierarchical representations over unaided linear prose.

Recent advancements in large language models (LLMs)  \cite{openai2023gpt4, team2023gemini, anthropic2024claude} have offered new avenues for relieving the burden of paper reading. LLM-based chatbots, for instance, can quickly generate abstract-length overviews or summaries of papers \cite{lenharo2024chatgpt}, often employing strategies like long context windows or retrieval-augmented generation (RAG) \cite{lewis2020retrieval, karpukhin2020dense} to handle lengthy documents.
While these tools can be useful for initial summarization, they often fall short in supporting deeper, section-specific understanding.
Users may find it difficult to obtain accurate answers for questions pertaining to specific sections, frequently needing to copy-paste paragraphs, and the generated responses can sometimes include fabricated references  \cite{Walters:2023aa} or errors in bibliographic details \cite{info:doi/10.2196/53164}. 
Furthermore, these flat summaries typically omit the paper's inherent section structure, thereby eliminating crucial navigational cues that users rely on for verification and further in-depth analysis.

Researchers in human–computer interaction (HCI) and machine learning (ML) have begun to address these shortcomings.
Systems like Sensecape \cite{sensecape} facilitate multilevel abstraction and fluid transitions in complex information tasks, while ScholarPhi \cite{scholarphi2020} augments the local comprehension of scientific PDFs with just‑in‑time definitions of terms and symbols. 
LLM-based methods, such as RAPTOR for multi‑level summary trees for information retrieval \cite{sarthi2024raptorrecursiveabstractiveprocessing}, and CHANGES for mapping papers into hierarchical structures to highlight key arguments and their contrasts for better summarization\cite{zhang2023contrastivehierarchicaldiscoursegraph}, further attest to the value of hierarchical structure.  However, a gap remains: these powerful algorithmic techniques have not yet been fully translated into user-centric interfaces that seamlessly unite hierarchical navigation with verifiable, LLM-generated summaries at multiple granularities.

To bridge this gap, we introduce \sysname{}, a novel language model-augmented paper reader designed to enhance the efficiency and effectiveness of academic reading. \sysname{} transforms a traditional linear paper into an interactive, hierarchical tree structure. Each section and subsection is initially represented by a concise, LLM-generated summary, with an option to reveal the full underlying content on demand. This design empowers users to: (i) Quickly grasp the core ideas of a paper by reading the key points for each section and paragraph;
(ii) Selectively and recursively explore sections of interest, focusing on information most relevant to their goals;
(iii) Verify information easily, as every summary node is linked to its dedicated source within the original text.

By integrating hierarchical summarization with an interactive tree view, \sysname{} transforms dense articles into navigable, verifiable maps of ideas. We demonstrate through a within-subject user study comparing \sysname{} with a standard PDF reader that our approach improves users' efficiency and effectiveness in skimming and goal-directed reading tasks. Finally, \sysname{} is publicly available as a Chrome extension\footnote{ \url{https://chromewebstore.google.com/detail/treereader/nhgffkcciononplndadobbkoomkjknkm}}, and we plan to collect additional feedback from real-world users to guide future iterations and improvements.
\section{Related work}
Scientific reading is an inherently goal-driven activity, yet existing AI-powered tools often provide limited support for how researchers engage with academic texts. Many current interfaces do not adequately reflect the hierarchical nature of scholarly writing, forcing users to mentally reconstruct this structure. In this section, we explore the landscape of related work across three key areas and highlight the need for a unified system that bridges advanced LLM capabilities with a user-centric, structure-aware design.

\subsection{AI-Assisted Reading Tools}

A primary focus of AI-assisted reading tools has been to enhance local comprehension and streamline aspects of the reading process. For instance, tools like ScholarPhi provide just-in-time definitions of terms \cite{head2021scholarphi}, while Papeos overlays talk videos to explain dense content \cite{kim2023papeos}. While these systems support understanding at the sentence or paragraph level, they typically do not surface the global structure necessary for synthesis or comparison across a document.

Other tools are geared towards literature discovery and managing citation context. including CiteSee \cite{chang2023citesee}, Relatedly \cite{palani2023relatedly}, and ComLittee \cite{kang2023comlittee}, which help researchers find relevant works or organize related papers. While powerful for the foraging stage of research, these tools offer minimal support once a user begins to read a specific paper in-depth. Similarly, systems like Scim \cite{fok2023scim}, and PaperWave \cite{yahagi2024paperwave} aim to assist skimming by selecting salient passages or translating papers into audio, often lack fine-grained user control over summary detail and may not provide robust mechanisms for verifying summarized content against the original source.

Recent research has also investigated how device interfaces and content rendering impact reading behaviour. Studies highlighting the differences between deep and skim reading across devices \cite{chen2023deep} and the cognitive burden of interface overload, such as from excessive browser tabs \cite{hwong2021tab}, emphasize the need for cognitively supportive reading interfaces. However, even systems that restructure content layout, like SideNoter \cite{abekawa2016sidenoter} for PDF annotation and browsing, or AI-Resilient Text Rendering \cite{li2024airesilient}, which adapts text presentation, often fall short of simultaneously supporting high-level navigation and deep, verifiable comprehension. In contrast, \sysname{} is designed to seamlessly support both skimming and deep reading of a single paper, adapting to varied user goals while prioritizing reliability and traceability through its hierarchical summarization and source-linking.

\subsection{Hierarchical and Cognitive Visualization Interfaces}
The value of non-linear representations in improving text comprehension and recall is well-established in cognitive and educational research. Studies consistently show that non-linear text structures enhance performance, especially when users need to navigate complex information or synthesize content across different sections \cite{sharifah2010effects, nonLinear}. These findings resonate strongly with sensemaking frameworks in HCI, which characterize reading as an active process of constructing hierarchical mental representations \cite{pirolli2005sensemaking}.

Informed by this, the HCI community has been developing AI-assisted tools that support ideation and synthesis, with a notable shift towards systems that scaffold human sensemaking rather than fully automating it \cite{ye2025designspacerecentaiassisted}. Several tools exemplify this by enabling users to externalize cognitive structures and support hierarchical reasoning. Scholastic \cite{scholastic} introduces hierarchical clustering into interpretive text analysis, helping users identify and navigate emergent themes through interactive visualizations. Building further on structural representation, IdeaSynth \cite{IdeaSynth} visualizes research ideas as modular, connected nodes, enabling users to iteratively develop and refine their ideas within a spatial and hierarchical canvas. Furthering this mixed-initiative approach, ScholarMate enables researchers to arrange text snippets on a non-linear canvas, leveraging AI for theme suggestion and multi-level summarization while ensuring human oversight and traceability to aid complex sensemaking tasks \cite{scholarmate}.

Other tools have specifically focused on structuring literature synthesis across multiple documents. Synergi \cite{synergi} and Threddy \cite{threddy} empower researchers to organize and interactively explore academic content by creating personalized hierarchical structures of literature. While these systems are powerful for synthesis across multiple sources, \sysname{}'s primary focus is on enhancing the navigation and comprehension within a single, often complex, academic paper.

\section{Formative study}
\begin{figure}
    \centering
\includegraphics[width=0.8\linewidth]{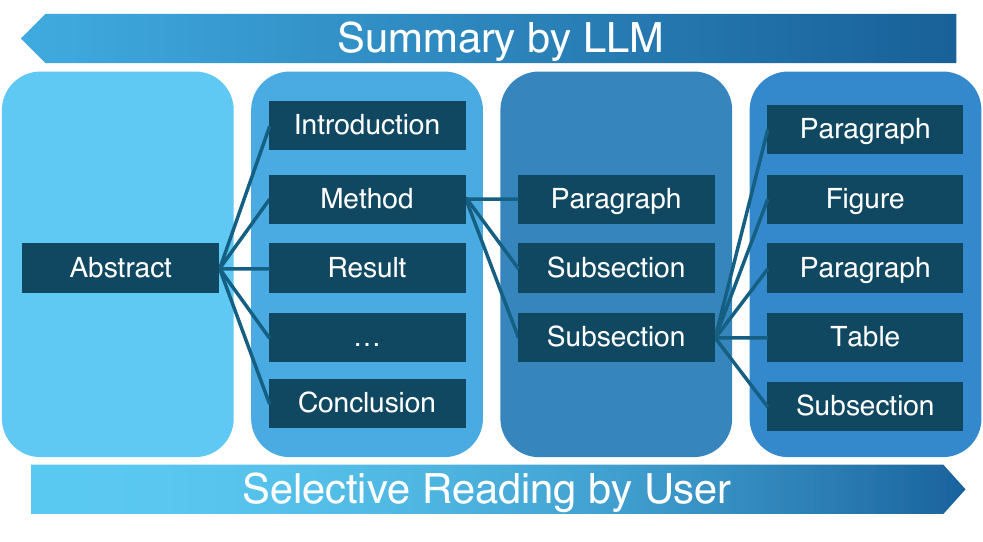}
    \caption{Section tree. \sysname{} uses the section tree as the structural backbone for presenting a paper's content. Each node in the tree represents a paragraph, table, figure, or section, with the contents of a section as child nodes. \sysname{} displays only the summaries of its children when displaying a section, which reduces the amount of texts the user need to read for targeting the useful child nodes.}
    \label{fig:section-tree}
\end{figure}

To capture a diverse range of perspectives and ensure the generalizability of our tool design, we interviewed participants from various academic disciplines, including chemistry, artificial intelligence, and human-computer interaction. Our goal was to understand three key dimensions of academic paper reading: \textbf{(1)} the challenges researchers encounter when engaging with papers in their field, \textbf{(2)} how they currently use LLM-based tools to support their reading and comprehension, and \textbf{(3)} their perceptions of the reliability and potential roles of LLMs in academic practices, such as peer review, literature review synthesis and research grant review. These insights informed our system design by highlighting critical user pain points and expectations around AI-assisted reading.

We conducted semi-structured interviews with six graduate researchers (5 men, 1 woman). The interviews were structured around open-ended questions aligned with our three dimensions of interest: reading challenges, use of LLM-based tools, and attitudes toward LLMs in high-stakes context. To ground the discussion, we used peer review as a concrete example when exploring participants’ opinions.
\\\textbf{C1: Information Targeting.} A primary challenge identified by participants was the difficulty in efficiently locating the most relevant and novel information within academic papers. All six interviewees expressed frustration with the need to sift through extensive content to extract key insights. They noted that much of a paper often reiterates familiar background or less critical details, while the core contributions are sparsely distributed throughout the text.
One participant explained, \textit{``There is too much information that is not interesting, and the interesting content is distributed over 10 pages.''}
%Another participant remarked the reason: \textit{``Most papers are incremental. If the authors focus on their key contribution, they will not have much to write.''}
\\\textbf{C2: Information Summary.} Reading summaries was a common strategy participants used to improve efficiency. Many participants expressed dissatisfaction with author-written abstracts and conclusions, stating that these often missed the key information needed to understand the actual contribution. Nearly all participants ($n=5$) reported using LLM-based tools to generate concise summaries, with three mentioning NotebookLM\footnote{\url{https://notebooklm.google/}} for audio-based comprehension. However, they noted that such tools were mainly useful for papers outside their core field, as the generated summaries were often too verbose for familiar topics.
\\\textbf{C3: LLM Reliability in High-Stakes Contexts.} As noted earlier, nearly all participants reported using LLM-based tools for paper comprehension, particularly for summarization.
However, when asked about the potential use of LLMs in peer review, a context where reliability is important, all participants expressed varying degrees of skepticism regarding the quality of information provided by these models.
Most participants believed that LLMs could provide limited support in reviewing manuscripts, such as identifying superficial issues like formatting or grammar.  
However, a majority ($n=4$) raised concerns that using LLMs for peer reviews might degrade their overall quality. In contrast, two participants saw potential for LLMs to complement human reviewers by providing more objective assessments. 
This mixed perspective on the role of LLM-based tools in peer review indicates that LLMs' reliability is still a significant concern, preventing their widespread adoption in high-stakes tasks like peer review.

%As one participant put it, “LLMs could be more objective, and sometimes human reviewers are not objective enough.”

%Three participants shared instances where LLMs hallucinated while explaining papers they had read. 
%However, when asked about using LLMs to review manuscripts, surprisingly, many participants mentioned that it has been a problem that human reviewers give low-quality review comments and the introduction of LLMs in the review process may provide a hope to improve it. 

%some participants disagreed that the comments from LLMS can be trusted without human review

%, while some participants also believe adding reviews from LLMs can make the review more objective.

\subsection{Design goals}

To ensure our system meaningfully addresses the real-world needs of academic users, we established three corresponding design goals to guide the development of our system:

\textbf{DG1: Enable hierarchical information exploration.}
To support the users targeting the information they need (\textbf{C1}), the system should present information hierarchically, allowing users to efficiently filter, navigate, and prioritize content. Users should be able to quickly grasp the structure of a paper and choose to expand or collapse sections of content based on relevance, thereby reducing the cognitive load associated with deciding what to read in detail.

\textbf{DG2: Provide summaries at multiple levels.}
To address the difficulty participants faced in information summaries (\textbf{C2}), the system should provide summaries not only for the entire paper but also for individual sections and paragraphs. This will allow users to access the right level of detail depending on their needs, whether they are skimming quickly or seeking a specific part. These summaries also serve as cues to help users focus on key content, which complements the goal of helping users find information efficiently (\textbf{C1}).

\textbf{DG3: Enhance reliability of LLM outputs.}
Given participants' concerns about the reliability of LLM-generated content (\textbf{C3}), the system should avoid using LLMs to the scope where their knowledge does not cover. The system should also incorporate mechanisms to help the user verify the output generated by LLMs.
%This includes grounding generated summaries closely to the original text, providing citations or references back to source passages.

\section{Design of \sysname{}}

Building upon these design goals, we designed \sysname{}, a paper reader which presents the contents of papers in a hierarchical way (See Figure~\ref{fig:system}), following the section tree structure of the paper (See Figure~\ref{fig:section-tree}). 
When displaying a section, instead of immediately displaying the original text of paragraphs and subsections, \sysname{} initially presents LLM-generated summaries that wrap the underlying content. Users can then choose to reveal finer details via the user interface as needed.
This approach significantly reduces the volume of information shown at each level, helping users navigate papers more efficiently and focus their attention on the most relevant parts.

\subsection{User interface \& Features}

The user interface of \sysname{} is organized into three columns as described below.

\textbf{Navigation Tree (Left Column):} 
This column presents a traditional section tree view of the paper. 
Users can select a specific node to focus on, expand or collapse parent nodes to view their children. 
Besides quick navigation, this column is designed to help the user have a sense of the location of the current node they are reading.

\textbf{Main contents (Middle Column):} 
The middle column displays the content of the section the user is currently viewing.
The nodes (including paragraphs, sections, figures, and tables) appear as cards arranged linearly in a column, which the user can scroll through.
For paragraph and section nodes, a summary of key points is shown for quick reading. In this way, we achieve \textbf{DG2} and provide a summary at multiple levels to the user.

Based on the information presented in the view, the user can adjust the level of detail they wish to view.
At the bottom of the card list, a navigation button (``←'') allows the user to return to the parent section if they choose not to explore further details.
For section nodes that have child nodes, a navigation button (``→'') is available, letting the user view those child nodes to access the full content beyond the summaries.
By selectively entering the view of the sections, the user only reads the information that is important to them and reads other information only in the form of a high-level summary. In this way, we enable a hierarchical information exploration and achieve \textbf{DG1}.

\textbf{Contextual Information (Right Column):} 
The rightmost column displays contextual information for the currently selected node. The top portion shows all figures from the node’s sub-tree, while the bottom portion presents either the original paragraph text (for paragraph nodes) or the titles and summaries of each sub-section and paragraph (for section nodes).

\subsection{LLM-based hierarchical summary}

As revealed in our formative study, concerns about LLM reliability present a significant barrier to their use in assisting with paper reading.
To address this issue and achieve \textbf{DG3}, we focus on two key characteristics of LLMs: (1) LLMs often lack access to the most up-to-date knowledge; (2) LLMs tend to overlook details when processing long inputs.

These observations guided our design decisions. First, we constrain the role of the LLM to information summarization only, avoiding reliance on potentially outdated or inaccurate domain knowledge. Second, we implement a recursive summarization approach to manage input length: rather than summarizing all content at once, the LLM summarizes the summaries of child nodes at each hierarchical level. This significantly reduces input size, especially for top-level sections.

Additionally, we instruct the LLM to attach source references to each key point in the summary. The source reference is displayed to the user when the user hovers the cursor on the end of each key point, enabling users to easily review the original content and identify potential errors.

\subsection{Other implementation Details}

\sysname{} is built with the \texttt{React} framework and uses OpenAI's \texttt{GPT-4o} to process the HTML extracted from Springer Nature publications \footnote{\url{https://www.springernature.com/}}.

\section{User evaluation}

We conducted a within-subject, semi-structured study to evaluate the effectiveness of \sysname{} compared to a standard PDF reader. Five graduate-level Computer Science researchers (3 men, 1 woman, 1 non-binary) participated in the 80-minute study. None of the participants had taken part in the earlier formative study. Each participant was compensated \$20. The study protocol was approved by the institution’s research ethics board.

Each participant completed two conditions: using a PDF Reader (C1) and using \sysname{} (C2), to read two different scientific papers (P1, P2) in counterbalanced order. For each paper, participants had up to 30 minutes total: 5 minutes for skimming and up to 25 minutes for deep reading.

We designed two comprehension tasks to reflect common academic reading practices: skimming for quick insight and deep reading for detailed understanding. After each condition, participants completed self-report questionnaires and preference interviews comparing both tools.

\textbf{Skimming Activity}:
This task assessed participants' ability to quickly locate key information. Skimming, defined as sacrificing depth for speed~\cite{just1987psychology}, was tested with a 5-minute reading period followed by multiple-choice and short-answer questions (Table \ref{tab:selfreport}). The questions targeted high-level understanding and the identification of central contributions.

\textbf{Deep Reading Activity}:
Following the skimming task, participants engaged in a deeper reading session of up to 25 minutes. They answered 8 open-ended questions (Appendix.\ref{sec:raw-responses}) designed to require close reading and information retrieval. All questions were shown upfront, and both accuracy and time to completion were recorded.

\textbf{Cognitive Load}:
After participants finished reading each paper, they completed the standard NASA Task Load Index (NASA-TLX) (Table \ref{tab:nasa}), which we used to measure cognitive load following the use of both tools for [P1, P2].
\section{Results}

We analyzed participants' experiences with \sysname{} along three key dimensions: its support for \textbf{efficient skimming}, its effectiveness in \textbf{deep reading} tasks, and its impact on \textbf{cognitive load}. While the study involved a small number of participants, their qualitative feedback revealed promising directions for improving and refining \sysname{}.

\textbf{\sysname{} shows potential to support efficient skimming.} Participants expressed positive initial reactions toward \sysname{}, highlighting its usefulness for quickly reviewing papers. One participant remarked, \textit{"I would definitely use \sysname{} every day for my initial literature review (P3)."} Others noted that the system's concise summaries and structured navigation made it easier to identify relevant sections, particularly in unfamiliar topics. These findings suggest that \sysname{} could streamline literature review workflows by helping users grasp objectives, structure, and key contributions more efficiently.

\begin{figure}
    \centering
    \includegraphics[width=0.8\linewidth]{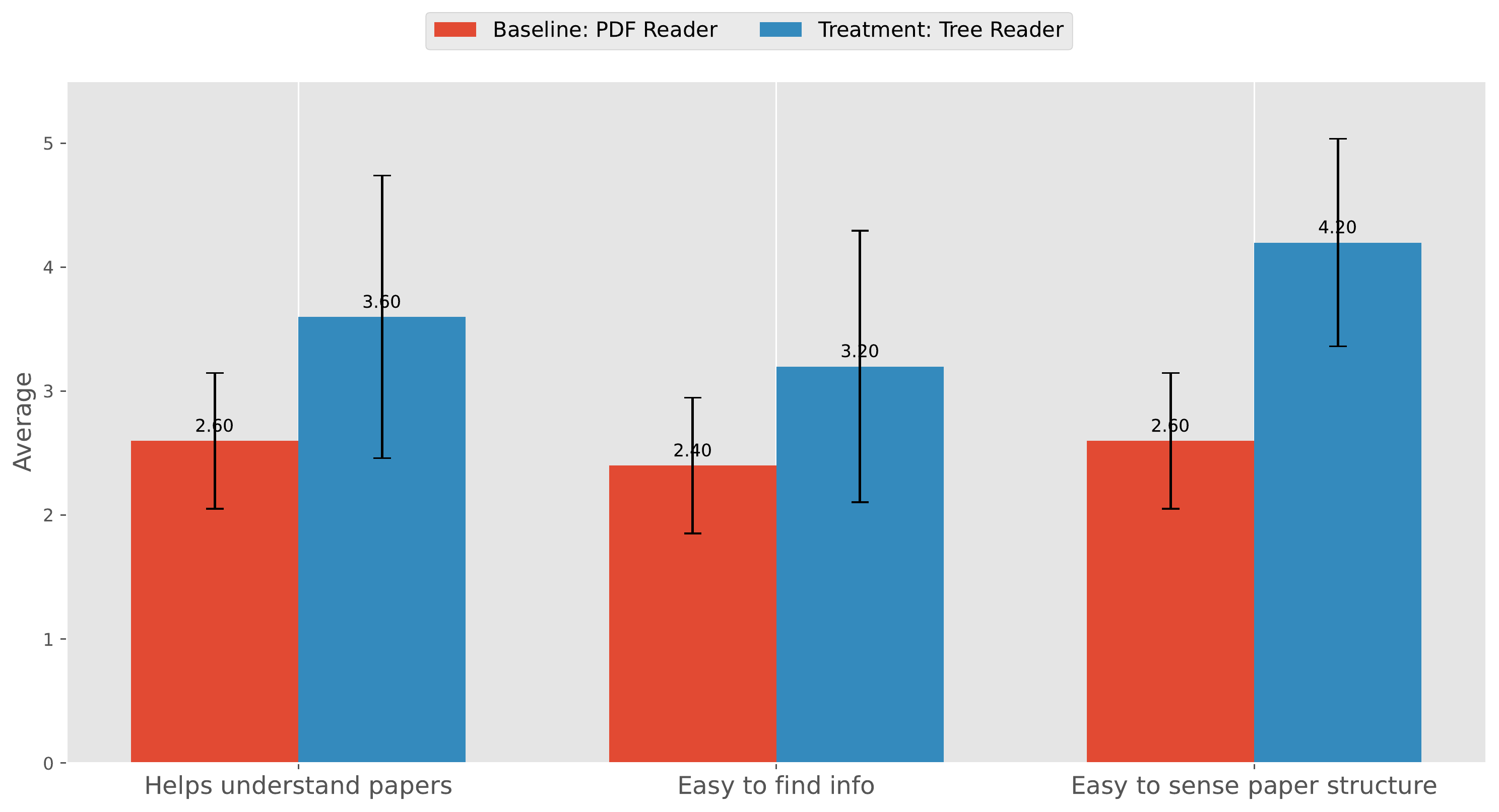}
    \caption{User feedback. After the deep reading activity, participants were asked whether (i) they understand the paper efficiently, (ii) it is easy to find information, (iii) it is easy to make sense of a section before looking at the details. On average, participants gave more positive responses to all three aspects when using \sysname{}. Please see Figure \ref{fig:postactivity-bar}, \ref{fig:postactivity-stack} for more details.
    \label{fig:post}}
    \vspace{-1em}
\end{figure}

\textbf{Participants had mixed experiences with deep reading tasks.} As shown in Figure \ref{fig:post}, using \sysname{} improved deep reading performance on average. However, individual responses varied: some found that \sysname{} aided sense-making and navigation, while others encountered difficulty retrieving specific details. These perspectives indicate that while \sysname{} supports structured exploration, it may benefit from refinements for in-depth reading. Despite this, several participants suggested enhancements like keyword-based search and dynamic subtree generation, reflecting enthusiasm for \sysname{}'s interactive potential.

\textbf{\sysname{} may help reduce cognitive load over time.} Participants reported that \sysname{} felt less cognitively demanding than traditional PDF readers once they became accustomed to its structure (Figure \ref{fig:nasa-bar}, \ref{fig:nasa-stack}). Although some experienced initial unfamiliarity, they acknowledged that continued use could reduce effort when engaging with complex academic texts.

\section{Conclusion}

We introduced \sysname{}, a novel tree-structured paper reader designed to enhance the efficiency of academic reading. Users can navigate through a hierarchical structure of sections using buttons, guided by LLM-generated summaries provided at both the section and paragraph levels.
By presenting summaries within a tree structure, \sysname{} reduces the amount of text users need to read, while also helping them locate relevant information more effectively through informative preview cues.
Finally, we conducted a user study to evaluate how \sysname{} impacts users’ performance in both skimming and deep reading tasks. The results provide evidence that \sysname{} can reduce cognitive load when reading long review papers.

This work has several limitations. Firstly, user responses may have been influenced by the novelty effect or unfamiliarity with \sysname{} compared to their extensive experience with PDF readers. Second, the study excluded common interaction features such as \texttt{Ctrl+F}, and omitted many real-world functionalities, potentially limiting the assessment of \sysname{}'s practical utility; a field study could address this and the Hawthorne effect. Lastly, the small and homogeneous group of five participants constrains the generalizability of our findings across diverse academic disciplines, reading habits, and LLM tool usage patterns.

\section*{Acknowledgments}
A.A.-G. thanks Anders G. Frøseth for his generous support. A.A.-G. also acknowledges the generous support of Natural Resources Canada and the Canada 150 Research Chairs program. This research is part of the University of Toronto’s Acceleration Consortium, which receives funding from the Canada First Research Excellence Fund (CFREF).
We also acknowledge the support of the Natural Sciences and Engineering Research Council of Canada (NSERC), [funding reference number RGPIN-2024-04348 and RGPIN-2024-06005].

\clearpage

%%%

{
\small
\bibliography{main}

\begin{thebibliography}{37}
\providecommand{\natexlab}[1]{#1}
\providecommand{\url}[1]{\texttt{#1}}
\expandafter\ifx\csname urlstyle\endcsname\relax
  \providecommand{\doi}[1]{doi: #1}\else
  \providecommand{\doi}{doi: \begingroup \urlstyle{rm}\Url}\fi

\bibitem[Abdelshaheed(2023)]{doi:10.1177/21582440231191795}
Bothina S.~M. Abdelshaheed.
\newblock Student versus expert outlines in reading comprehension: The effect of collaborative construction.
\newblock \emph{SAGE Open}, 13\penalty0 (3):\penalty0 21582440231191795, 2023.
\newblock \doi{10.1177/21582440231191795}.
\newblock \url{https://doi.org/10.1177/21582440231191795}.

\bibitem[Abekawa and Aizawa(2016)]{abekawa2016sidenoter}
Takeshi Abekawa and Akiko Aizawa.
\newblock Sidenoter: Scholarly paper browsing system based on pdf restructuring and text annotation.
\newblock In \emph{Proceedings of the 26th International Conference on Computational Linguistics}, pages 1--10. ACL, 2016.

\bibitem[Anthropic(2024)]{anthropic2024claude}
Anthropic.
\newblock The claude 3 model family: Opus, sonnet, haiku.
\newblock 2024.
\newblock \url{https://www-cdn.anthropic.com/de8ba9b01c9ab7cbabf5c33b80b7bbc618857627/Model_Card_Claude_3.pdf}.

\bibitem[Chang et~al.(2023)Chang, Zhang, Bragg, Head, Lo, Downey, and Weld]{chang2023citesee}
Joseph~Chee Chang, Amy~X. Zhang, Jonathan Bragg, Andrew Head, Kyle Lo, Doug Downey, and Daniel~S. Weld.
\newblock Citesee: Augmenting citations in scientific papers with persistent and personalized historical context.
\newblock In \emph{Proceedings of the 2023 CHI Conference on Human Factors in Computing Systems}, pages 1--15. ACM, 2023.
\newblock \doi{10.1145/3544548.3580847}.

\bibitem[Chelli et~al.(2024)Chelli, Descamps, Lavou{\'e}, Trojani, Azar, Deckert, Raynier, Clowez, Boileau, and Ruetsch-Chelli]{info:doi/10.2196/53164}
Mika{\"e}l Chelli, Jules Descamps, Vincent Lavou{\'e}, Christophe Trojani, Michel Azar, Marcel Deckert, Jean-Luc Raynier, Gilles Clowez, Pascal Boileau, and Caroline Ruetsch-Chelli.
\newblock Hallucination rates and reference accuracy of chatgpt and bard for systematic reviews: Comparative analysis.
\newblock \emph{J Med Internet Res}, 26:\penalty0 e53164, May 2024.
\newblock ISSN 1438-8871.
\newblock \doi{10.2196/53164}.
\newblock \url{https://www.jmir.org/2024/1/e53164}.

\bibitem[Chen et~al.(2023)Chen, Srivastava, Jain, and Marlow]{chen2023deep}
Xiuge Chen, Namrata Srivastava, Rajiv Jain, and Jennifer Marlow.
\newblock Characteristics of deep and skim reading on smartphones vs. desktop: A comparative study.
\newblock In \emph{Proceedings of the 2023 CHI Conference on Human Factors in Computing Systems}, pages 1--13. ACM, 2023.
\newblock \doi{10.1145/3544548.3581174}.

\bibitem[Fok et~al.(2023)Fok, Kambhamettu, Soldaini, Bragg, Lo, Head, Hearst, and Weld]{fok2023scim}
Raymond Fok, Hita Kambhamettu, Luca Soldaini, Jonathan Bragg, Kyle Lo, Andrew Head, Marti~A. Hearst, and Daniel~S. Weld.
\newblock Scim: Intelligent skimming support for scientific papers.
\newblock In \emph{Proceedings of the 28th International Conference on Intelligent User Interfaces (IUI '23)}, pages 476--490, New York, NY, USA, 2023. Association for Computing Machinery.
\newblock ISBN 979-8-4007-0106-1.
\newblock \doi{10.1145/3581641.3584034}.
\newblock \url{https://doi.org/10.1145/3581641.3584034}.

\bibitem[Head et~al.(2021{\natexlab{a}})Head, Lo, Kang, Fok, Skjonsberg, Weld, and Hearst]{head2021scholarphi}
Andrew Head, Kyle Lo, Dongyeop Kang, Raymond Fok, Sam Skjonsberg, Daniel~S. Weld, and Marti~A. Hearst.
\newblock Scholarphi: Augmenting scientific papers with just-in-time, position-sensitive definitions of terms and symbols.
\newblock In \emph{Proceedings of the 2021 CHI Conference on Human Factors in Computing Systems}, pages 1--12. ACM, 2021{\natexlab{a}}.
\newblock \doi{10.1145/3411764.3445648}.

\bibitem[Head et~al.(2021{\natexlab{b}})Head, Lo, Kang, Fok, Skjonsberg, Weld, and Hearst]{scholarphi2020}
Andrew Head, Kyle Lo, Dongyeop Kang, Raymond Fok, Sam Skjonsberg, Daniel~S. Weld, and Marti~A. Hearst.
\newblock Augmenting scientific papers with just-in-time, position-sensitive definitions of terms and symbols.
\newblock In \emph{Proceedings of the 2021 CHI Conference on Human Factors in Computing Systems}, CHI '21, New York, NY, USA, 2021{\natexlab{b}}. Association for Computing Machinery.
\newblock ISBN 9781450380966.
\newblock \doi{10.1145/3411764.3445648}.
\newblock \url{https://doi.org/10.1145/3411764.3445648}.

\bibitem[Hong et~al.(2022)Hong, Marsh, Feuston, Ruppert, Brubaker, and Szafir]{scholastic}
Matt-Heun Hong, Lauren~A. Marsh, Jessica~L. Feuston, Janet Ruppert, Jed~R. Brubaker, and Danielle~Albers Szafir.
\newblock Scholastic: Graphical human-ai collaboration for inductive and interpretive text analysis.
\newblock In \emph{Proceedings of the 35th Annual ACM Symposium on User Interface Software and Technology}, pages 1--12. Association for Computing Machinery, 2022.
\newblock \doi{10.1145/3526113.3545681}.
\newblock \url{https://dl.acm.org/doi/10.1145/3526113.3545681}.

\bibitem[Hwong et~al.(2021)Hwong, Klein, Henley, and Kittur]{hwong2021tab}
John Hwong, Benjamin~A. Klein, Austin~Z. Henley, and Aniket Kittur.
\newblock When the tab comes due: Challenges in the cost structure of browser tab usage.
\newblock In \emph{Proceedings of the 2021 CHI Conference on Human Factors in Computing Systems}, pages 1--14. ACM, 2021.
\newblock \doi{10.1145/3411764.3445585}.

\bibitem[Just and Carpenter(1987)]{just1987psychology}
Marcel~Adam Just and Patricia~Ann Carpenter.
\newblock \emph{The psychology of reading and language comprehension.}
\newblock Allyn \& Bacon, 1987.

\bibitem[Kang et~al.(2022)Kang, Chang, Kim, and Kittur]{threddy}
Hyeonsu Kang, Joseph~Chee Chang, Yongsung Kim, and Aniket Kittur.
\newblock Threddy: An interactive system for personalized thread-based exploration and organization of scientific literature.
\newblock In \emph{Proceedings of the 35th Annual ACM Symposium on User Interface Software and Technology}, pages 1--15. Association for Computing Machinery, 2022.
\newblock \doi{10.1145/3526113.3545660}.
\newblock \url{https://dl.acm.org/doi/10.1145/3526113.3545660}.

\bibitem[Kang et~al.(2023{\natexlab{a}})Kang, Soliman, Latzke, Chang, and Bragg]{kang2023comlittee}
Hyeonsu~B. Kang, Nouran Soliman, Matt Latzke, Joseph~Chee Chang, and Jonathan Bragg.
\newblock Comlittee: Literature discovery with personal elected author committees.
\newblock In \emph{Proceedings of the 2023 CHI Conference on Human Factors in Computing Systems}, pages 1--13. ACM, 2023{\natexlab{a}}.
\newblock \doi{10.1145/3544548.3581264}.

\bibitem[Kang et~al.(2023{\natexlab{b}})Kang, Wu, Chang, and Kittur]{synergi}
Hyeonsu~B Kang, Tongshuang Wu, Joseph~Chee Chang, and Aniket Kittur.
\newblock Synergi: A mixed-initiative system for scholarly synthesis and sensemaking.
\newblock In \emph{Proceedings of the 36th Annual ACM Symposium on User Interface Software and Technology}, pages 1--19. Association for Computing Machinery, 2023{\natexlab{b}}.
\newblock \doi{10.1145/3586183.3606759}.
\newblock \url{https://dl.acm.org/doi/10.1145/3586183.3606759}.

\bibitem[Karpukhin et~al.(2020)Karpukhin, O{\u{g}}uz, Min, Lewis, Wu, Edunov, Chen, and Yih]{karpukhin2020dense}
Vladimir Karpukhin, Barlas O{\u{g}}uz, Sewon Min, Patrick Lewis, Ledell Wu, Sergey Edunov, Danqi Chen, and Wen-tau Yih.
\newblock Dense passage retrieval for open-domain question answering.
\newblock \emph{arXiv preprint arXiv:2004.04906}, 2020.

\bibitem[Kim et~al.(2023)Kim, Latzke, Bragg, Zhang, and Chang]{kim2023papeos}
Tae~Soo Kim, Matt Latzke, Jonathan Bragg, Amy~X. Zhang, and Joseph~Chee Chang.
\newblock Papeos: Augmenting research papers with talk videos.
\newblock In \emph{Proceedings of the 36th Annual ACM Symposium on User Interface Software and Technology}, pages 1--12. ACM, 2023.
\newblock \doi{10.1145/3586183.3606770}.

\bibitem[Lenharo(2024)]{lenharo2024chatgpt}
Mariana Lenharo.
\newblock Chatgpt turns two: How the ai chatbot has changed scientists’ lives.
\newblock \emph{Nature}, 636\penalty0 (8042):\penalty0 281--282, 2024.

\bibitem[Lewis et~al.(2020)Lewis, Perez, Piktus, Petroni, Karpukhin, Goyal, K{\"u}ttler, Lewis, Yih, Rockt{\"a}schel, et~al.]{lewis2020retrieval}
Patrick Lewis, Ethan Perez, Aleksandra Piktus, Fabio Petroni, Vladimir Karpukhin, Naman Goyal, Heinrich K{\"u}ttler, Mike Lewis, Wen-tau Yih, Tim Rockt{\"a}schel, et~al.
\newblock Retrieval-augmented generation for knowledge-intensive nlp tasks.
\newblock \emph{Advances in Neural Information Processing Systems}, 33:\penalty0 9459--9474, 2020.

\bibitem[Li et~al.(2024)Li, Liang, Peng, and Yin]{li2024airesilient}
Zhuoyan Li, Chen Liang, Jing Peng, and Ming Yin.
\newblock An ai-resilient text rendering technique for reading and skimming documents.
\newblock In \emph{Proceedings of the 2024 CHI Conference on Human Factors in Computing Systems}, pages 1--12. ACM, 2024.
\newblock \doi{10.1145/3613904.3642699}.

\bibitem[McEneaney(1994)]{nonLinear}
John~E. McEneaney.
\newblock Cognitive processing of hyperdocuments: When does nonlinearity help?
\newblock In \emph{Proceedings of the SIGCHI Conference on Human Factors in Computing Systems}, pages 57--63, 1994.
\newblock \doi{10.1145/168466.168508}.
\newblock \url{https://dl.acm.org/doi/10.1145/168466.168508}.

\bibitem[McKoon(1977)]{McKoon1977OrganizationOI}
Gail McKoon.
\newblock Organization of information in text memory.
\newblock \emph{Journal of Verbal Learning and Verbal Behavior}, 16:\penalty0 247--260, 1977.
\newblock \url{https://api.semanticscholar.org/CorpusID:144071803}.

\bibitem[Miller(1956)]{miller1956magical}
George~A Miller.
\newblock The magical number seven, plus or minus two: Some limits on our capacity for processing information.
\newblock \emph{Psychological review}, 63\penalty0 (2):\penalty0 81, 1956.

\bibitem[Muttalib(2010)]{sharifah2010effects}
Sharifah Amani~Abdul Muttalib.
\newblock The effects of linear and non-linear text on students’ performance in reading, 2010.

\bibitem[OpenAI(2023)]{openai2023gpt4}
OpenAI.
\newblock Gpt-4 technical report, 2023.

\bibitem[Palani et~al.(2023)Palani, Naik, Downey, Zhang, Bragg, and Chang]{palani2023relatedly}
Srishti Palani, Aakanksha Naik, Doug Downey, Amy~X. Zhang, Jonathan Bragg, and Joseph~Chee Chang.
\newblock Relatedly: Scaffolding literature reviews with existing related work sections.
\newblock In \emph{Proceedings of the 2023 CHI Conference on Human Factors in Computing Systems}, pages 1--14. ACM, 2023.
\newblock \doi{10.1145/3544548.3580954}.

\bibitem[Pirolli and Card(2005)]{pirolli2005sensemaking}
Peter Pirolli and Stuart Card.
\newblock The sensemaking process and leverage points for analyst technology as identified through cognitive task analysis.
\newblock In \emph{Proceedings of international conference on intelligence analysis}, volume~5, pages 2--4. McLean, VA, USA, 2005.

\bibitem[Pu et~al.(2024)Pu, Feng, Grossman, Hope, Mishra, Latzke, Bragg, Chang, and Siangliulue]{IdeaSynth}
Kevin Pu, K.~J.~Kevin Feng, Tovi Grossman, Tom Hope, Bhavana~Dalvi Mishra, Matt Latzke, Jonathan Bragg, Joseph~Chee Chang, and Pao Siangliulue.
\newblock Ideasynth: Iterative research idea development through evolving and composing idea facets with literature-grounded feedback, 2024.
\newblock \url{https://arxiv.org/abs/2410.04025}.

\bibitem[Sarthi et~al.(2024)Sarthi, Abdullah, Tuli, Khanna, Goldie, and Manning]{sarthi2024raptorrecursiveabstractiveprocessing}
Parth Sarthi, Salman Abdullah, Aditi Tuli, Shubh Khanna, Anna Goldie, and Christopher~D. Manning.
\newblock Raptor: Recursive abstractive processing for tree-organized retrieval, 2024.
\newblock \url{https://arxiv.org/abs/2401.18059}.

\bibitem[Suh et~al.(2023)Suh, Min, Palani, and Xia]{sensecape}
Sangho Suh, Bryan Min, Srishti Palani, and Haijun Xia.
\newblock Sensecape: Enabling multilevel exploration and sensemaking with large language models.
\newblock In \emph{Proceedings of the 36th Annual ACM Symposium on User Interface Software and Technology}, UIST '23, New York, NY, USA, 2023. Association for Computing Machinery.
\newblock ISBN 9798400701320.
\newblock \doi{10.1145/3586183.3606756}.
\newblock \url{https://doi.org/10.1145/3586183.3606756}.

\bibitem[Team et~al.(2023)Team, Anil, Borgeaud, Wu, Alayrac, Yu, Soricut, Schalkwyk, Dai, Hauth, et~al.]{team2023gemini}
Gemini Team, Rohan Anil, Sebastian Borgeaud, Yonghui Wu, Jean-Baptiste Alayrac, Jiahui Yu, Radu Soricut, Johan Schalkwyk, Andrew~M Dai, Anja Hauth, et~al.
\newblock Gemini: a family of highly capable multimodal models.
\newblock \emph{arXiv preprint arXiv:2312.11805}, 2023.

\bibitem[Thalmann et~al.(2019)Thalmann, Souza, and Oberauer]{Thalmann:2019aa}
Mirko Thalmann, Alessandra~S Souza, and Klaus Oberauer.
\newblock How does chunking help working memory?
\newblock \emph{J Exp Psychol Learn Mem Cogn}, 45\penalty0 (1):\penalty0 37--55, Jan 2019.
\newblock ISSN 1939-1285 (Electronic); 0278-7393 (Linking).
\newblock \doi{10.1037/xlm0000578}.

\bibitem[Walters and Wilder(2023)]{Walters:2023aa}
William~H. Walters and Esther~Isabelle Wilder.
\newblock Fabrication and errors in the bibliographic citations generated by chatgpt.
\newblock \emph{Scientific Reports}, 13\penalty0 (1):\penalty0 14045, 2023.
\newblock \doi{10.1038/s41598-023-41032-5}.
\newblock \url{https://doi.org/10.1038/s41598-023-41032-5}.

\bibitem[Yahagi et~al.(2024)Yahagi, Chujo, Harada, Han, Sugiyama, and Naemura]{yahagi2024paperwave}
Yuchi Yahagi, Rintaro Chujo, Yuga Harada, Changyo Han, Kohei Sugiyama, and Takeshi Naemura.
\newblock Paperwave: Listening to research papers as conversational podcasts scripted by llm.
\newblock \emph{arXiv preprint arXiv:2410.15023}, 2024.

\bibitem[Ye et~al.(2025{\natexlab{a}})Ye, Lee, Varona, Huang, and Nobre]{scholarmate}
Runlong Ye, Patrick Yung~Kang Lee, Matthew Varona, Oliver Huang, and Carolina Nobre.
\newblock Scholarmate: A mixed-initiative tool for qualitative knowledge work and information sensemaking, 2025{\natexlab{a}}.
\newblock \url{https://arxiv.org/abs/2504.14406}.

\bibitem[Ye et~al.(2025{\natexlab{b}})Ye, Varona, Huang, Lee, Liut, and Nobre]{ye2025designspacerecentaiassisted}
Runlong Ye, Matthew Varona, Oliver Huang, Patrick Yung~Kang Lee, Michael Liut, and Carolina Nobre.
\newblock The design space of recent ai-assisted research tools for ideation, sensemaking, and scientific creativity, 2025{\natexlab{b}}.
\newblock \url{https://arxiv.org/abs/2502.16291}.

\bibitem[Zhang et~al.(2023)Zhang, Liu, and Zhang]{zhang2023contrastivehierarchicaldiscoursegraph}
Haopeng Zhang, Xiao Liu, and Jiawei Zhang.
\newblock Contrastive hierarchical discourse graph for scientific document summarization, 2023.
\newblock \url{https://arxiv.org/abs/2306.00177}.

\end{thebibliography}
\bibliographystyle{assets/plainnat}
}

%%%

\clearpage

\appendix
% \appendixtoc

%%%

\lstdefinestyle{base}{
  breaklines=true,
  basicstyle=\fontsize{10}{10}\selectfont\ttfamily
}
\newtcolorbox{demobox}[2]{breakable,colback=#2!5!white,colframe=#2!75!black,fonttitle=\bfseries,title=#1}
\definecolor{teal}{RGB}{50, 220, 220} % teal

\section{Prompts to language models}

When processing a paper into a tree, we use the following prompt to convert the paragraphs in the tree into key points and the corresponding evidence. In the prompts for generating the key points of a section, we provide the key points of the children nodes of the section as the input to the prompt with some minor adjustments to the prompt.

\begin{demobox}{Prompts for summarizing paragraphs}{teal}
\begin{lstlisting}[style=base]
Here is an abstract of a scientific paper and a specific paragraph from the same paper. Please read both and then summarize the paragraph in the context of the abstract. 
<Abstract>
{abstract}
</Abstract>
<Paragraph>
{node.content}
</Paragraph>
<Requirement>
You are required to output a summary of the paragraph in the format of 2~5 key points. The key points should not be more than 70 words in total. The key points should summary the original content comprehensively.

Return your summary in with a JSON with a single key "points", whose value is a list with 2~5 JSON objects with the following keys:
"point" (str): A key point of the paragraph. The key point should be a complete sentence stating an important facts. 
"evidence" (str): A copy of the original text that support the point.
</Requirement>
\end{lstlisting}
\end{demobox}

\section{Detail of Responses to Post-Skimming self-reported questions}

For each paper, participants were given 5 minutes to skim the paper using the assigned tool, followed by five self-reported questions (Table~\ref{tab:selfreport}). We have provided participants' responses to these questions in this section.

\textit{Note that when we analyzed the data, we mapped Strongly disagreement to 1, Disagree to 2, Neutral to 3, Agree to 4, and Strongly agree to 5.}

\begin{table}[H]
\small
\renewcommand{\arraystretch}{1.5}
\setlength{\tabcolsep}{4pt}
\caption{Survey questions on review paper understanding after five-minute skimming.}
\label{tab:selfreport}
\begin{tabularx}{\textwidth}{|X|c|c|c|c|c|}
\hline
\textbf{Question/Response} & \textbf{Strongly disagree} & \textbf{Disagree} & \textbf{Neutral} & \textbf{Agree} & \textbf{Strongly agree} \\
\hline
I felt like that I understood the objective of this review paper. & & & & & \\
\hline
I felt like that I understood how the review paper was organized. & & & & & \\
\hline
I felt like that I understood the key messages of this review paper. & & & & & \\
\hline
I felt like that I understood the strengths and weaknesses of this review paper. & & & & & \\
\hline
I felt like that I understood the challenges in the field this review paper is talking about. & & & & & \\
\hline
\end{tabularx}
\end{table}

\begin{figure}[H]
    \centering
    \includegraphics[width=1\linewidth]{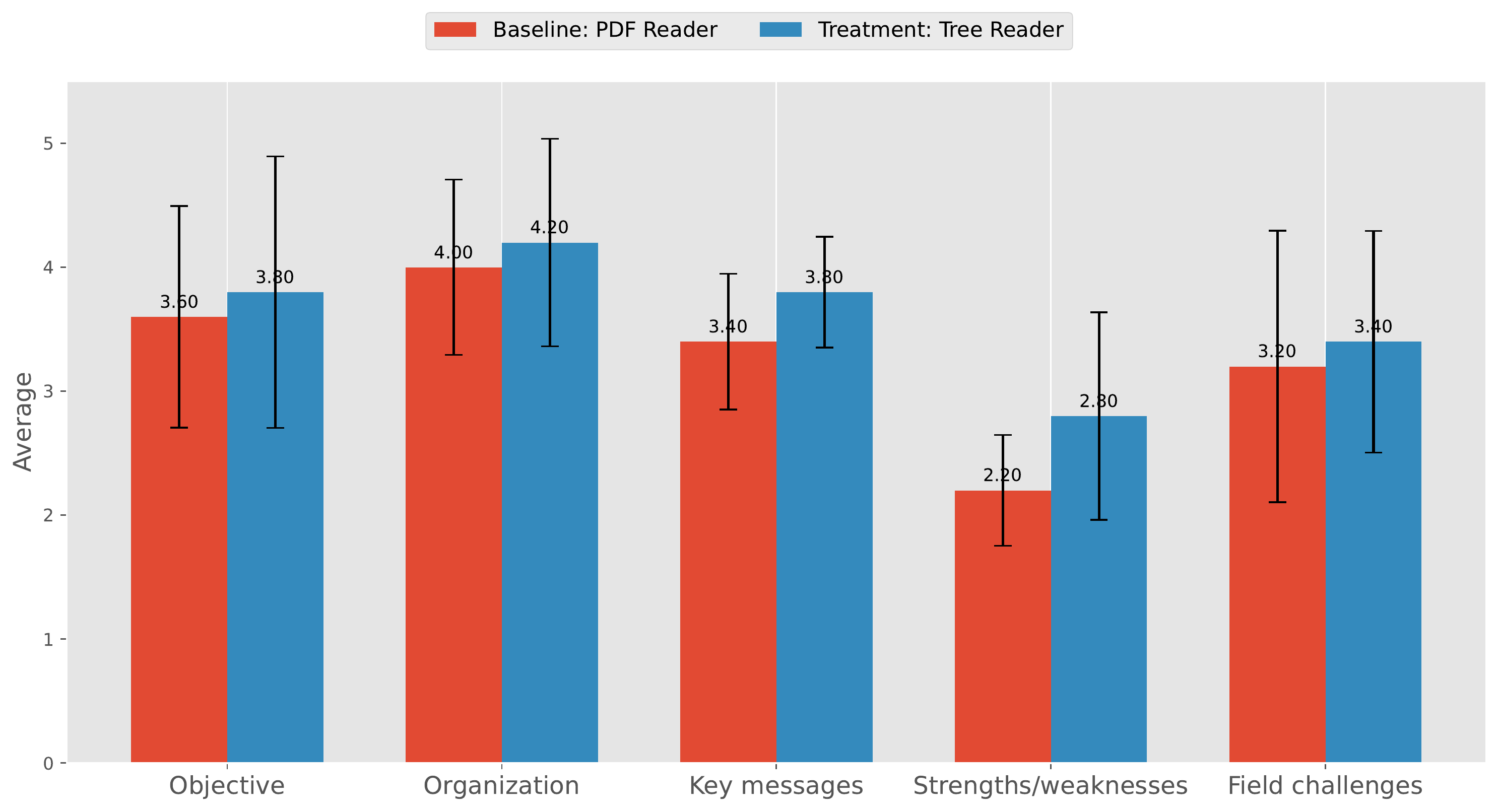}
    \caption{The average responses from participants on the post-skimming questions. Participants used both PDF Reader and \sysname{}, and this figure shows that on average participants had a higher score on every post-skimming question with \sysname{}.
    \label{fig:selfreport-bar}}
    \vspace{-1em}
\end{figure}

\begin{figure}[H]
    \centering
    \includegraphics[width=1\linewidth]{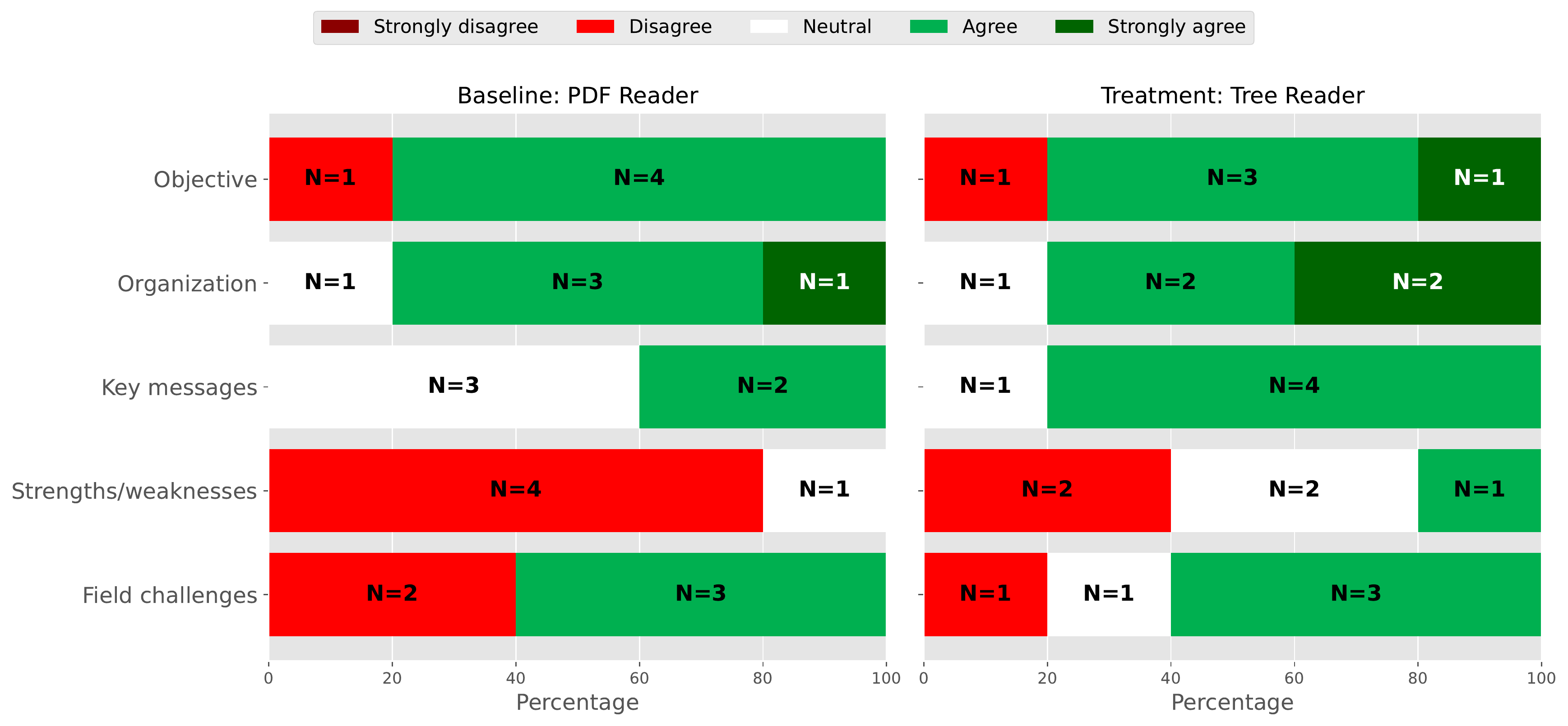}
    \caption{The side-by-side comparison between PDF Reader and \sysname{}, we reported the number of each response for each post-skimming question.
    \label{fig:selfreport-bar}}
    \vspace{-1em}
\end{figure}

\section{Detail of Responses to Post-Activity self-reported questions}

For each paper, after five-minute skimming skimming and at most 25 minutes deep reading, participants were asked five self-reported questions regarding their understandings of the paper and the overall experience with \sysname{} (Table~\ref{tab:post-deepreading}). We have provided participants' responses to these questions in this section.

\textit{Note that when we analyzed the data, we mapped Strongly disagreement to 1, Disagree to 2, Neutral to 3, Agree to 4, and Strongly agree to 5.}

\begin{table}[H]
\small
\renewcommand{\arraystretch}{1.5}
\setlength{\tabcolsep}{4pt}
\caption{Survey questions on review paper understanding after both skimming and deep reading activities}
\label{tab:post-deepreading}
\begin{tabularx}{\textwidth}{|X|c|c|c|c|c|}
\hline
\textbf{Question/Response}                                                                                    & Strongly disagree & Disagree & Neutral & Agree & Strongly agree \\ \hline
I am confident about my answers to Q1 - Q4.                                                                   &                   &          &         &       &                \\ \hline
I am confident about my answers to Q5 - Q8.                                                                   &                   &          &         &       &                \\ \hline
This tool is very reliable. I can count on it to be correct all the time                                      &                   &          &         &       &                \\ \hline
It is easy for me to find the information I need                                                              &                   &          &         &       &                \\ \hline
It is easy to sense making a sections of the scientific papers before looking at the details of this sections &                   &          &         &       &                \\ \hline
\end{tabularx}
\end{table}

\begin{figure}[H]
    \centering
    \includegraphics[width=1\linewidth]{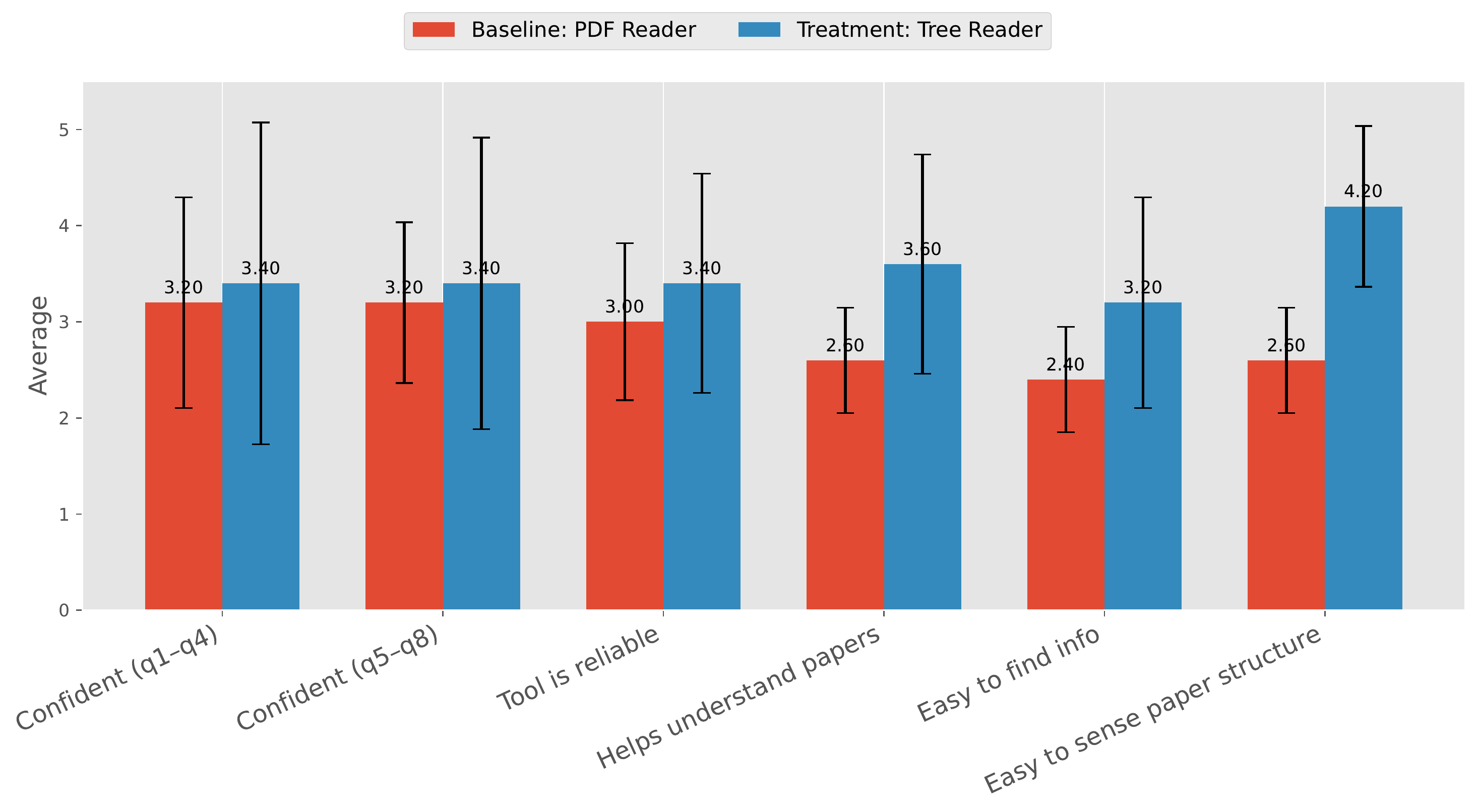}
    \caption{The average responses from participants on the post-activity questions. Participants used both PDF Reader and \sysname{}, and this figure shows that on average participants had a higher score on every post-activity question with \sysname{}, especially when it is about the difficulty to sense and understand paper structure and find information.
    \label{fig:postactivity-bar}}
    \vspace{-1em}
\end{figure}

\begin{figure}[H]
    \centering
    \includegraphics[width=1\linewidth]{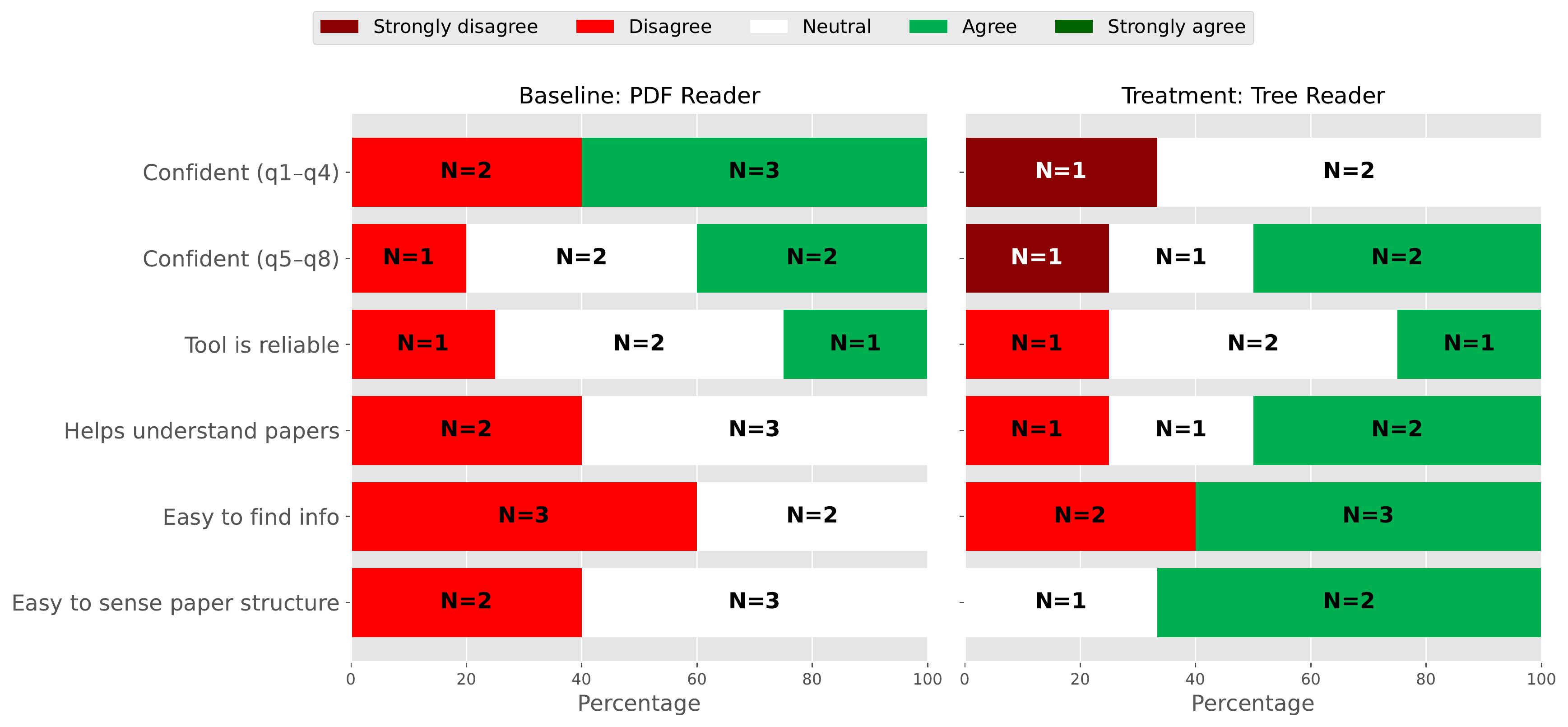}
    \caption{The side-by-side comparison between PDF Reader and \sysname{}, we reported the number of each response for each post-activity question.
    \label{fig:postactivity-stack}}
    \vspace{-1em}
\end{figure}

\section{Detail of Responses to the NASA Task Load Index (TLX) questions}

For each paper, at the very end, after the post-activity questions, we showed the participants six questions from the NASA Task Load Index (TLX) (Table~\ref{tab:nasa}). We have provided participants' responses to these questions in this section.

\textit{Note that when we analyzed the data, we mapped Very Low to 1, Below Average to 2, Average to 3, Above Average to 4, and Very High to 5.}

\begin{table}[H]
\small
\renewcommand{\arraystretch}{1.5}
\setlength{\tabcolsep}{4pt}
\caption{Survey questions on the NASA Task Load Index (TLX). Note that except for \textbf{Performance}, a lower score indicates a lower cognitive load.}
\label{tab:nasa}
\begin{tabularx}{\textwidth}{|X|c|c|c|c|c|}
\hline
\textbf{Question/Response}                                                                                         & Very Low & Below Average & Average & Above Average & Very High \\ \hline
Mental Demand: How mentally demanding was the task?                                                                &          &               &         &               &           \\ \hline
Physical Demand: How physically demanding was the task?                                                            &          &               &         &               &           \\ \hline
Temporal Demand: How hurried or rushed was the pace of the task?                                                   &          &               &         &               &           \\ \hline
Performance: How would you rate your success in accomplishing the assigned task?                                   &          &               &         &               &           \\ \hline
Effort: How hard did you have to work to accomplish your level of performance?                                     &          &               &         &               &           \\ \hline
Frustration: How insecure, discouraged, irritated, stressed, and annoyed\textless{}/strong\textgreater am are you? &          &               &         &               &           \\ \hline
\end{tabularx}
\end{table}

\begin{figure}[H]
    \centering
    \includegraphics[width=1\linewidth]{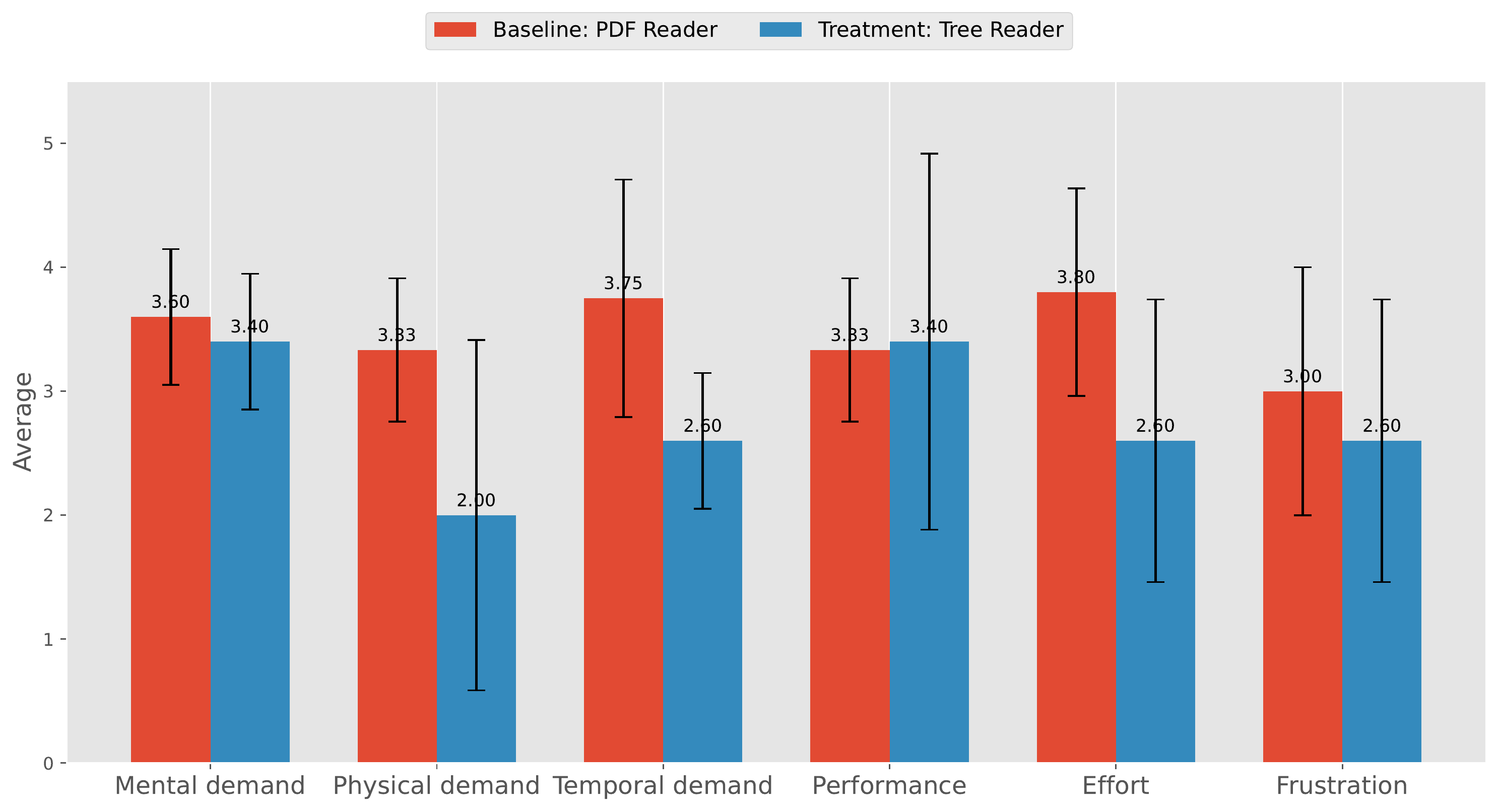}
    \caption{The average responses from participants on the NASA TLX questions. Participants used both PDF Reader and \sysname{}, and this figure shows that on average participants had a lower cognitive load on every post-activity question with \sysname{}. However, despite the improvements by \sysname{}, participants did not rate their performance with \sysname{} higher than PDF Reader. We acknowledge this limitation might be because of the small sample size and the differences between the two papers. We encourage future work with \sysname{} to be deployed with a large sample size and more diverse papers.
    \label{fig:nasa-bar}}
    \vspace{-1em}
\end{figure}

\begin{figure}[H]
    \centering
    \includegraphics[width=1\linewidth]{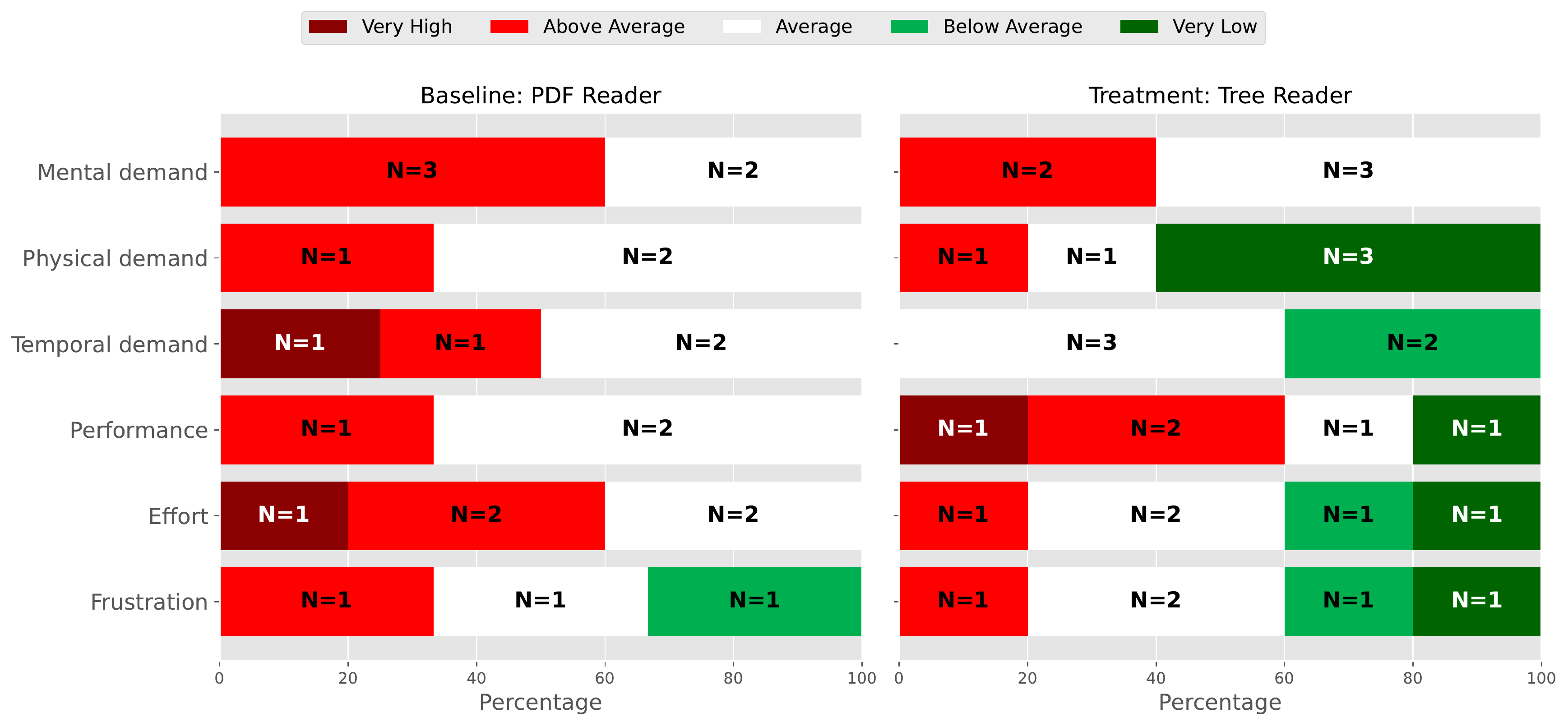}
    \caption{The side-by-side comparison between PDF Reader and \sysname{}, we reported the number of each response for each NASA TLX question.
    \label{fig:nasa-stack}}
    \vspace{-1em}
\end{figure}

\section{Participants' raw responses to the tasks while doing the deep reading}
\label{sec:raw-responses}

The two papers used in the user evaluation are 1. \textit{Towards trustworthy LLMs: a review on debiasing and dehallucinating in large language models}\footnote{\url{https://link.springer.com/article/10.1007/s10462-024-10896-y}} and 2. \textit{Artificial intelligence for literature reviews: opportunities and challenges}\footnote{\url{https://link.springer.com/article/10.1007/s10462-024-10902-3}}.

In the following, we use normal text to represent the responses made by using PDF Reader and mark the responses made with \sysname{} \textcolor{mattergreen1}{green}.
 
 \subsection{Paper 1: Towards trustworthy LLMs: a review on debiasing and dehallucinating in large language models} 
\textbf{Q1: What is bias in LLMs}
\begin{itemize}
\item \textcolor{black}{P1: Bias in LLMs refers to behaviours that reinforce undesirable, offensive, or stereotypical content from the training corpora.}
\item \textcolor{black}{P2: the process of detecting, mitigating, or eliminating biases, especially in NLP and machine learning, ensuring that models and algorithms neither inherit nor propagate unequal, unfair or unsuitable information}
\item \textcolor{mattergreen1}{P3: Bias in LLms are caused due to training data.They are categorized into racial, gender, and political biases}
\item \textcolor{black}{P4: LLMs inherit toxic, offensive, misleading, stereotypical,
and other behaviors that are harmful or discriminatory}
\item \textcolor{mattergreen1}{P5: toxic, offensive, misleading, stereotypical, and other behaviors that are harmful or discriminatory}
\end{itemize} 
\textbf{Q2: What are the major causes of bias in LLMs?}
\begin{itemize}
\item \textcolor{black}{P1: 2. The main cause seems to be inherited bias from vast training corpora.}
\item \textcolor{black}{P2: Racial and religious biases, Gender and orientation biases \& Political and cultural biases}
\item \textcolor{mattergreen1}{P3: The training data, they often reflect stereotypes for example: associating cooking with women and CEOs with men}
\item \textcolor{black}{P4: 1. bias in the data leads to LLM being biased
2. If the training process doesn't debias the LLM from learning these bad artifacts in the data
3. If no procedure is employed for guardrail LLMs from generating biased contnet}
\item \textcolor{mattergreen1}{P5: trained on vast amounts of data, which are often sourced from vast and diverse online corpora}
\end{itemize} 
\textbf{Q3: What are the major causes of hallucinations in LLMs?}
\begin{itemize}
\item \textcolor{black}{P1: 3. One cause is that training corpora can contain incorrect information or overemphasize certain information. It can also be caused by the model itself depending on its architecture or algorithm.}
\item \textcolor{black}{P2: Data level \& Model level}
\item \textcolor{mattergreen1}{P3: Flawed training data, Weaker model architectures, Exposure bias}
\item \textcolor{black}{P4: The model architecture, if too weak
The sampling/decoding algorithm and the introduction of too much noise
The exposure of models to biased data during training}
\item \textcolor{mattergreen1}{P5: lack of knowledge or erroneous information.}
\end{itemize} 
\textbf{Q4: Why evaluating bias is considered easier than evaluating hallucination?}
\begin{itemize}
\item \textcolor{black}{P1: Bias can be straightforwardly identified (for example, through word correlations), but hallucinations are much harder to detect because the claim of the LLM must be evaluated on its own.}
\item \textcolor{black}{P2: With human involvement,
human annotators tend to look for shortcuts to make the task easier, so they are more
likely to base their judgments on surface attributes such as fluency and language complexity, rather than expending more effort on detecting authenticity.}
\item \textcolor{mattergreen1}{P3: Traditional metrics like BLEU and ROUGE are inadequate for evaluating hallucination in generated text. Whereas we have metrics such as Word Embedding Association Test (WEAT) and Sentence Encoder Association Test (SEAT) can be used to evaluating bias.}
\item \textcolor{black}{P4: Because people expect current models to be general purposed, which imposed harder metrics for evaluation.
Traditiona evaluation metrics, especially the statistics based like ROUGE, often has little correlation with the actual amount of hallucination. And it's also the case for model based evaluation. Thus, most of the work require human based eval.
By comparison, bias evaluation can involve using embedding models / statistical models, which are human free.}
\item \textcolor{mattergreen1}{P5: Evaluating bias is more systematic than evaluating hallucination. biases can be analyzed from embedding level and word level, while hallucination has to be put under context to understand, and sometime require human labeling.}
\end{itemize} 
\textbf{Q5: What does counterfactual data augmentation do as a debias method}
\begin{itemize}
\item \textcolor{black}{P1: CDA involves “injecting” training data that contrasts with the biases of the existing data, for example by changing sentences to have feminine pronouns.}
\item \textcolor{black}{P2: biases in pre-trained language models largely arise from imbalances
in their training data. A direct approach to counter these biases involves rebalancing the training data. Counterfactual data augmentation (CDA) (Zhao et al. 2018) is a primary method for data rebalancing, which is widely used (Zmigrod et al. 2019; Webster et al. 2020; Barikeri et al. 2021). To mitigate gender bias between male and female demographic groups, it is essential to ensure that gender-neutral terms exhibit consistent relationships with gender-specific terms.}
\item \textcolor{mattergreen1}{P3: Counterfactual data augmentation (CDA) is a primary method for rebalancing training data. CDA helps mitigate gender bias by ensuring gender-neutral terms relate equally to both genders}
\item \textcolor{black}{P4: TO generate synthetic data and balance the distribution of different types, ie, debiasing}
\item \textcolor{mattergreen1}{P5: generate more data to ensure gender-neutral terms relate equally to both genders}
\end{itemize} 
\textbf{Q6: What type of bias is probed by the WinoBias dataset?}
\begin{itemize}
\item \textcolor{black}{P1: WinoBias focuses on gender bias with respect to occupations.}
\item \textcolor{black}{P2: gender bias}
\item \textcolor{mattergreen1}{P3: WinoBias is a dataset designed to investigate gender bias in coreference resolution.}
\item \textcolor{black}{P4: gender bias,}
\item \textcolor{mattergreen1}{P5: gender bias}
\end{itemize} 
\textbf{Q7: What challenges do traditional metrics like BLEU, ROUGE, and METEOR face when evaluating hallucination?}
\begin{itemize}
\item \textcolor{black}{P1: Because they use n-gram similarity, they are not sensitive to the level of hallucination.}
\item \textcolor{black}{P2: these metrics, which rely on n-gram to quantify the similarity between generated text and reference text, face challenges in evaluating the level of hallucination (Dhingra et al. 2019; Durmus et al. 2020).}
\item \textcolor{mattergreen1}{P3: Conventional metrics like ROUGE and BLEU show low correlation with human judgment for evaluating hallucinations in generated content. The PARENT metric, which uses reference text, aligns more closely with human judgment than traditional target text metrics}
\item \textcolor{black}{P4: It has challenge in indicating the level of hallucination, and low correlation with human evaluation (ground truth)}
\item \textcolor{mattergreen1}{P5: they rely on n-gram to quantify the similarity between generated text and reference text, face challenges in evaluating the level of hallucination}
\end{itemize} 
\textbf{Q8: Regarding the challenge of dehallucination, what limits human feedback-based methods.}
\begin{itemize}
\item \textcolor{black}{P1: Human feedback can be unreliable due to subjectivity and human error.}
\item \textcolor{black}{P2: With human involvement, human annotators tend to look for shortcuts to make the task easier, so they are more
likely to base their judgments on surface attributes such as fluency and language complexity, rather than expending more effort on detecting authenticity}
\item \textcolor{mattergreen1}{P3: The human evaluation is accurate, it is labor-intensive and lacks reproducibility.}
\item \textcolor{black}{P4: human feedback can be subjective and unreliable due to personal biases and error
annotations, making it hard fully represent authenticity.}
\item \textcolor{mattergreen1}{P5: human annotators tend to look for shortcuts to make the task easier, so they are more likely to base their judgments on surface attributes such as fluency and language complexity, rather than expending more effort on detecting authenticity.}
\end{itemize} \subsection{Paper 2: Artificial intelligence for literature reviews: opportunities and challenges} 
\textbf{Q1: What are the phases/stages that a systematic literature review has?}
\begin{itemize}
\item \textcolor{mattergreen1}{P1: From the source content: "(i) Planning, (ii) Search, (iii) Screening, (iv) Data Extraction and Synthesis, (v) Quality Assessment, and (vi) Reporting."}
\item \textcolor{mattergreen1}{P2: (i) Planning, (ii) Search, (iii) Screening, (iv) Data Extraction and Synthesis, (v) Quality Assessment, and (vi) Reporting.}
\item \textcolor{black}{P3: (i) Planning, (ii) Search, (iii) Screening, (iv) Data Extraction and Synthesis, (v) Quality Assessment, and (vi) Reporting.}
\item \textcolor{mattergreen1}{P4: (i) Planning, (ii) Search, (iii) Screening, (iv) Data Extraction and Synthesis, (v) Quality Assessment, and (vi) Reporting.}
\item \textcolor{black}{P5: (i) Planning, (ii) Search, (iii) Screening, (iv) Data Extraction and Synthesis, (v)
Quality Assessment, and (vi) Reporting.}
\end{itemize} 
\textbf{Q2: Based on previous surveys/reviews, what are the main AI features in SLR tools?}
\begin{itemize}
\item \textcolor{mattergreen1}{P1: From the source text: "approach, text representation, human interaction, input, and output"}
\item \textcolor{mattergreen1}{P2: Approach' feature}
\item \textcolor{black}{P3: Approach,Text representation,Human interaction,Input,Output}
\item \textcolor{mattergreen1}{P4: screening and data extraction phases of SLRs}
\item \textcolor{black}{P5: Approach, text representation, human interaction, input, output}
\end{itemize} 
\textbf{Q3: What is the phase/stage that current AI tools help most in SLR?}
\begin{itemize}
\item \textcolor{mattergreen1}{P1: The primary focus of these tools seems to be in the screening phase, as well as data extraction.}
\item \textcolor{mattergreen1}{P2: DIDN’T ANSWER}
\item \textcolor{black}{P3: screening and data extraction phases}
\item \textcolor{mattergreen1}{P4: The screening phase}
\item \textcolor{black}{P5: screening}
\end{itemize} 
\textbf{Q4: Why do people do Systematic Literature Review?}
\begin{itemize}
\item \textcolor{mattergreen1}{P1: An SLR objectively summarizes all of the relevant literature for a particular research topic.}
\item \textcolor{mattergreen1}{P2: to understand pros and cons}
\item \textcolor{black}{P3: Its main goal is to meticulously identify and appraise all the relevant literature related to a specific research question, adhering to strict protocols to minimize biases.}
\item \textcolor{mattergreen1}{P4: The primary aim of SLRs is to identify and assess relevant literature while minimizing biases}
\item \textcolor{black}{P5: it helps researchers identify and appraise all the relevant literature related to a specific research question,}
\end{itemize} 
\textbf{Q5: What does quality assessment evaluate?}
\begin{itemize}
\item \textcolor{mattergreen1}{P1: The primary focus of these tools seems to be in the screening phase, as well as data extraction.}
\item \textcolor{mattergreen1}{P2: DIDN’T ANSWER}
\item \textcolor{black}{P3: The quality assessment phase evaluates the rigour and validity of the selected studies. This analysis provides evidence of the overall strength and the level of trustworthiness presented in the review.}
\item \textcolor{mattergreen1}{P4: performance, usability, and transparenc}
\item \textcolor{black}{P5: The quality assessment phase evaluates the rigour and validity of the selected
studies.}
\end{itemize} 
\textbf{Q6: What is the approach most adopted by the AI tools in the screening stage?}
\begin{itemize}
\item \textcolor{mattergreen1}{P1: AI is used to apply inclusion/exclusion criteria to papers. Usually, this involves a classifier trained on hand-picked examples with the hope of generalizing to a broader set of papers.}
\item \textcolor{mattergreen1}{P2: DIDN’T ANSWER}
\item \textcolor{black}{P3: It usually involves employing machine learning classifiers, which are trained on an initial set of user-selected papers and then used to identify additional relevant articles. This process frequently involves iteration, where the user refines the automatic classifications or selects new papers, followed by retraining the classifier to better identify further pertinent literatur}
\item \textcolor{mattergreen1}{P4: Machine learning classifiers}
\item \textcolor{black}{P5: Support Vector Machine (SVM)}
\end{itemize} 
\textbf{Q7: Regarding the validity of this survey, what type of validity is Snowballing method used to reduce?}
\begin{itemize}
\item \textcolor{mattergreen1}{P1: I couldn't find the name for the specific type of validity in the paper. However, snowballing was used in Semantic Scholar to validate the originally screened set of papers. Since the new process revealed no new relevant papers, this suggests the original methodology was comprehensive enough.}
\item \textcolor{mattergreen1}{P2: DIDN’T ANSWER}
\item \textcolor{black}{P3: To ensure that papers arent lost/ not incorporated. Snowballing helped to find more papers.}
\item \textcolor{mattergreen1}{P4: external validity}
\item \textcolor{black}{P5: internal validity}
\end{itemize} 
\textbf{Q8: Regarding the search engine tools, what are the bibliographic databases they use?}
\begin{itemize}
\item \textcolor{mattergreen1}{P1: They used Scopus, the SLR Toolbox, and CRAN.}
\item \textcolor{mattergreen1}{P2: Elicit, Consensus, and Perplexity use Semantic Scholar, while EvidenceHunt relies on PubMed}
\item \textcolor{black}{P3: Scopus and Dimensions, CORE TopAI Tools}
\item \textcolor{mattergreen1}{P4: Semantic Scholar, PubMed}
\item \textcolor{black}{P5: Semantic Scholar, PubMed, and a broader array of publishers, such as Wiley, Sage,
Europe PMC, Thieme, and Cambridge University Press}
\end{itemize}

%%%

\end{document}